\documentclass[10pt,nofootinbib,floatfix,a4,showpacs,
showkeys,notitlepage,twocolumn]{revtex4-1}

\pdfoutput=1

\usepackage{amsmath}
\usepackage{amssymb}
\usepackage{color}
\usepackage{graphicx}
\usepackage{hyperref}
\usepackage[utf8]{inputenc}
\usepackage[T1]{fontenc} 
\usepackage{slashed}
\allowdisplaybreaks

\def\nvl{n_{\text{VL}}}

\newcommand{\AddrLiege}{%
  IFPA, Dep. AGO, Universit\'e de Li\`ege, Bat B5, Sart-Tilman B-4000
  Li\`ege 1, Belgium.}
\newcommand{\AddrUdeA}{%
    Instituto de F\'{\i}sica, Universidad de Antioquia, A.A. 1226,
    Medellin, Colombia.}
\newcommand{\AddrVlc}{%
  Instituto de F\'{\i}sica Corpuscular
    (CSIC-Universitat de Val\`{e}ncia), Apdo. 22085, E-46071 Valencia,
    Spain.}
  \newcommand{\AddrRITS}{%
    Department of Theoretical Physics, School of Engineering Sciences,
    KTH Royal Institute of Technology, AlbaNova University Center, 106
    91 Stockholm, Sweden}

\begin{document}

\title{Diboson anomaly: heavy Higgs resonance and QCD vector-like exotics}

\author{D. Aristizabal Sierra}
\email{daristizabal@ulg.ac.be}
\affiliation{\AddrLiege}
\author{J. Herrero-Garcia}
\email{juhg@kth.se}
\affiliation{\AddrRITS}
\author{D. Restrepo}
\email{restrepo@udea.edu.co}
\affiliation{\AddrUdeA}
\author{A. Vicente}
\email{avelino.vicente@ific.uv.es}
\affiliation{\AddrVlc}

% $^a$\affiliation{\AddrLiege}
% $^b$\affiliation{\AddrVlc}
% $^c$\affiliation{\AddrUdeA}
%
\begin{abstract}
  The ATLAS collaboration (and also CMS) has recently reported an
  excess over Standard Model expectations for gauge boson pair
  production in the invariant mass region $1.8-2.2$ TeV. In the light
  of these results, we argue that such signal might be the first
  manifestation of the production and further decay of a heavy CP-even
  Higgs resulting from a type-I Two Higgs Doublet Model. We
  demonstrate that in the presence of colored vector-like fermions,
  its gluon fusion production cross-section is strongly enhanced, with
  the enhancement depending on the color representation of the new
  fermion states. Our findings show that barring the color triplet
  case, any QCD ``exotic'' representation can fit the ATLAS result in
  fairly large portions of the parameter space. We have found that if
  the diboson excess is confirmed and this mechanism is indeed
  responsible for it, then the LHC Run-2 should find: (i) a CP-odd
  scalar with mass below $\sim 2.3$ TeV, (ii) new colored states with
  masses below $\sim 2$ TeV, (iii) no statistically significant
  diboson events in the $W^\pm Z$ channel, (iv) events in the triboson
  channels $W^\pm W^\mp\,Z$ and $ZZZ$ with invariant mass amounting to
  the mass of the CP-odd scalar.
  % Its production cross-section at the LHC is dominated by the standard
  % gluon fusion process, strongly enhanced by the presence of new
  % vector-like colored fermions. Once produced, the heavy Higgs decays
  % to $WW$ and $ZZ$ final states, leading to the diboson excesses
  % observed. Regarding the vector-like states responsible for the heavy
  % Higgs production, we consider several $SU(3)_c$ representations:
  % ${\bf 3}$, ${\bf 6}$, ${\bf 8}$, ${\bf 10}$ and ${\bf 15}$. We show
  % that although a heavy scalar produced in gluon fusion via $SU(3)_c$
  % triplets cannot account for the diboson excess, larger $SU(3)_c$
  % multiplets can successfully account for the ATLAS diboson
  % anomaly. Taking into account current direct searches, we find that
  % colored sextets and octets are the best suited candidates and should
  % soon be produced at the run 2 of the LHC.
\end{abstract}
\maketitle

\section{Introduction}
\label{sec:intro}
The ATLAS collaboration has recently reported an excess over the
Standard Model (SM) expectations for gauge boson pair production in
the invariant mass region of $1.8-2.2$ TeV~\cite{Aad:2015owa}. The
statistical significance of the excess observed by ATLAS is $3.4
\sigma$, $2.6 \sigma$ and $2.9 \sigma$ in the $WZ$, $WW$ and $ZZ$
channels, respectively, although the hadronic nature of the search
makes it hard to distinguish gauge bosons implying some overlap
between these channels. The CMS collaboration has also reported some
moderate excesses in diboson searches both in hadronic
channels~\cite{Khachatryan:2014hpa} and in semileptonic
channels~\cite{Khachatryan:2014gha}, again at invariant masses around
$2.0$ TeV. Although the statistical significance is lower in this case
($1-2 \, \sigma$), the fact that these excesses occur at roughly the
same invariant mass value has made the \emph{diboson excess} a hot
subject in the community.

Although further data from the LHC Run-2 is required to confirm a
diboson overproduction at $1.8-2.2$ TeV, it is tempting to speculate
about new physics scenarios where these hints would be naturally
explained. The obvious explanation to the ATLAS and CMS hints is a
bosonic resonance that decays into a pair of SM gauge bosons. In order
to be able to explain the LHC data, the hypothetical candidate must
face two requirements~\cite{Aad:2015owa} (see also
\cite{Allanach:2015hba,Goncalves:2015yua,Bian:2015hda} for general
analyses of the ATLAS data): (i) it has to be produced with a
relatively high cross-section in the $\sim 1-10$ fb ballpark, and (ii)
it should be a narrow resonance (with $\Gamma \lesssim 200$ GeV)
decaying dominantly to a diboson final state. Many candidates with
these properties have been already proposed\footnote{Other recent
  works related to the diboson excess include an effective theory
  approach to the anomaly~\cite{Kim:2015vba} and implications on dark
  matter searches~\cite{Liew:2015osa}, grand
  unification~\cite{Bandyopadhyay:2015fka} or neutrinoless double beta
  decay and lepton flavor violation~\cite{Awasthi:2015ota}.}. Although
most references focus on spin-1 candidates, such as
$W^\prime/Z^\prime$ states in extended gauge
models~\cite{Fukano:2015hga,Hisano:2015gna,Franzosi:2015zra,Cheung:2015nha,Dobrescu:2015qna,Alves:2015mua,Gao:2015irw,Thamm:2015csa,Brehmer:2015cia,Cao:2015lia,Cacciapaglia:2015eea,Abe:2015jra,Heeck:2015qra,Allanach:2015hba,Abe:2015uaa,Carmona:2015xaa,Dobrescu:2015yba,Fukano:2015uga,Krauss:2015nba,Anchordoqui:2015uea,Bian:2015ota,Lane:2015fza,Low:2015uha,Terazawa:2015bsa,Arnan:2015csa,Niehoff:2015iaa,Dev:2015pga,Dobado:2015hha,Deppisch:2015cua,Aydemir:2015nfa,Dobado:2015nqa,Li:2015yya},
other alternatives are perfectly viable. Examples of such alternatives
are spin-2 states~\cite{Sanz:2015zha} and triboson production
\cite{Aguilar-Saavedra:2015rna}. Finally, spin-0 particles are among
the simplest candidates to explain the diboson excess and have been
considered in
\cite{Chiang:2015lqa,Cacciapaglia:2015nga,Sanz:2015zha,Chen:2015xql,Omura:2015nwa,Chao:2015eea,Arnan:2015csa,Petersson:2015rza,Zheng:2015dua,Fichet:2015yia,Dermisek:2015oja,Chen:2015cfa}.

In this paper we propose the heavy Higgs ($H$) of a type-I Two Higgs
Doublet Model (2HDM) as the resonance behind the diboson excess. Its
production cross-section at the LHC is dominated by the standard gluon
fusion process, strongly enhanced by the presence of new vector-like
(VL) colored fermions. Furthermore, a large production cross-section
for the heavy Higgs naturally implies negligible VL contributions to
the gluon fusion cross-section for the light Higgs ($h$), which
remains SM-like. Once produced, the heavy Higgs decays to $WW$ and $ZZ$
final states, leading to the diboson excesses found by ATLAS and
CMS. Regarding the VL states responsible for the heavy Higgs
production, we consider several $SU(3)_c$ representations: ${\bf 3}$,
${\bf 6}$, ${\bf 8}$, ${\bf 10}$ and ${\bf 15}$. QCD exotics
(non-triplet $SU(3)_c$ representations) are well motivated in this
context due to their large contributions to the gluon fusion
cross-section~\cite{Ilisie:2012cc}.

Another model with an extended scalar sector and VL colored states has
been put forward as an explanation to the diboson excess in
\cite{Chen:2015cfa}. Our setup differs from the one considered in this
reference in several aspects. First, in our case the heavy Higgs
resonance is embedded in a 2HDM framework which constrains its
couplings and decay modes. Second, we go beyond the fundamental
representation and explore the phenomenology induced by higher
$SU(3)_c$ multiplets. In fact, we will show that a heavy scalar
produced in gluon fusion driven by $SU(3)_c$ triplets cannot account
for the diboson excess. One needs the enhancement coming from a larger
color factor in order to achieve production cross-sections of the
required size. Finally, we will also comment on some technical
differences in the phenomenological analysis.

The rest of the manuscript is organized as follows. In
sec. \ref{sec:vl-quarks}, we discuss the features of extra VL colored
fermions, the structure of the scalar sector, the colored VL fermions
mass matrices and the relevant couplings. In sec. \ref{sec:pheno}, we
study the phenomenological aspects of our scenario, in particular we
discuss the different relevant aspects of the heavy Higgs production
cross-section: group theory factors, $\alpha_S$ RGE running and VL
fermion mass limits. In sec. \ref{sec:numerics}, we present our main
findings. Finally, in sec. \ref{sec:conclusions} we present our
conlcusions.
\section{Vector-like colored fermions}
\label{sec:vl-quarks}
The setup we consider involves extra VL colored fermions ($\nvl$ VL
generations) with the following $\left( SU(3)_c , SU(2)_L
\right)_{U(1)_Y}$ transformation properties:
\begin{align}
  \label{eq:new-vl-states}
  Q_L &= \left(d_R , {\bf 2} \right)_{\frac{1}{6}} \, ,
  &   Q_R &= \left(d_R , {\bf 2} \right)_{\frac{1}{6}} \, , 
  \nonumber\\
  U_L &= \left(d_R , {\bf 1} \right)_{\frac{2}{3}} \, ,
  &   U_R &= \left(d_R , {\bf 1} \right)_{\frac{2}{3}} \, , 
  \nonumber\\
  D_L &= \left(d_R , {\bf 1} \right)_{-\frac{1}{3}} \, , 
  & D_R &=
  \left(d_R , {\bf 1} \right)_{-\frac{1}{3}} \, ,
\end{align}
where $d_R=\boldsymbol{3}, \boldsymbol{6}, \boldsymbol{8},
\boldsymbol{10}$ and $\boldsymbol{15}$. We decompose the $SU(2)_L$ doublets
as
\begin{align}
\label{eq:doublet}
Q_{L,R} & =
\begin{pmatrix}
\widetilde U \\
\widetilde D 
\end{pmatrix}_{L,R}
\, ,
\end{align}
such that $Q\left(\widetilde U\right)=2/3$ and $Q\left(\widetilde
D\right)=-1/3$.  In addition to these states and the usual SM
fermions, the scalar sector involves two hypercharge $+1$ scalar
electroweak doublets, $H_1$ and $H_2$.

The resulting 2HDM must of course yield a scalar with SM-like
properties, something that in the absence of the VL states is readily
achievable by moving towards the decoupling
limit~\cite{Gunion:2002zf}. However, in the presence of the new VL
colored states this condition is not sufficient: the couplings to the
VL states can potentially modify the SM Higgs production
cross-section. Thus, in order to guarantee phenomenological
consistency we endow our setup with an additional $\mathbb{Z}_2$
symmetry under which
\begin{alignat}{2}
  \label{eq:transformations}
  X_\text{SM}&\to X_\text{SM}\, ,
  \qquad
  &Q&\to Q\, ,
  \nonumber\\
  (U,D)&\to -(U,D)\, ,
  \qquad
  &H_A&\to (-1)^A H_A\, .
\end{alignat} 
Under these transformations one is left with basically two
$\mathbb{Z}_2$-conserving sectors: one chiral (SM sector) and another
VL, each one with its own scalar doublet. Thus, in this sense the SM
and VL sectors are ``orthogonal'' up to scalar mixing, which can
always be taken such that both sectors are decoupled. Note that in
the absence of the VL fermions the resulting model would be a type-I
2HDM (see e.g. ref.~\cite{Branco:2011iw} for further details).

It is well known that chiral-VL quark mixing is subject to several
(stringent) constraints: In the up sector, these mixings induce
e.g. deviations in the $Zqq$ couplings which are severely
constrained. First and second generation mixing are constrained by
atomic parity violation experiments~\cite{Agashe:2014kda} and
measurements of $R_c=\Gamma(Z\to \bar cc)/\Gamma(Z\to \text{hadrons})$
at LEP~\cite{Agashe:2014kda,Aguilar-Saavedra:2013qpa}, and are by far
more stringent than those found for the third generation
\cite{AguilarSaavedra:2002kr}. In the down sector the most severe
constraints are derived from measurements of $R_b=\Gamma(Z\to \bar
bb)/\Gamma(Z\to \text{hadrons})$, and so are stronger for third
generation~\cite{AguilarSaavedra:2002kr}. First and second down-type
quark mixing are however severely constrained by meson mixing and
decays~\cite{Barenboim:2001fd}. In our setup, only color triplets
enable writing $\mathbb{Z}_2$-invariant renormalizable chiral-VL
mixing terms. Indeed, an appealing feature of QCD exotics
($d_R=\boldsymbol{6}, \boldsymbol{8}, \boldsymbol{10},
\boldsymbol{15}$) is the intrinsic absence of renormalizable-induced
chiral-VL mixing, assured by color invariance. Thus, in the triplet
case, and only in that case, one has to worry about the size of the
couplings controlling chiral-VL mixing. Two simple ways can be
envisaged, either their values are phenomenologically fixed (can be
fixed to zero) or the $\mathbb{Z}_2$ symmetry is promoted to a
$\mathbb{Z}_4$ symmetry under which the SM fields and $H_2$ are
neutral, while $Q \to -i \, Q$, $(U,D) \to i \, (U,D)$ and $H_1\to
-H_1$ \footnote{For chiral quarks in the color triplet representation,
  a setup resembling in some aspects this one has been considered in
  ref. \cite{Camargo:2015xca}}. In what follows whenever referring to
the color triplet we will assume the former.

Under the above working assumptions, the SM quarks combined with the
$H_2$ doublet induce the following $\mathbb{Z}_2$-invariant Yukawa
interactions
\begin{align}
  \label{eq:Yukawa-Z4-inv-SM}
  -\mathcal L_{Y}^{\text{SM}} &=
  \overline{q}_L\cdot \boldsymbol{h_u}\cdot u_R\,\widetilde{H}_2
  + \overline{q}_L\cdot \boldsymbol{h_d}\cdot d_R\,H_2 \nonumber \\
  & + \overline{\ell}_L\cdot \boldsymbol{h_e}\cdot e_R\,H_2
  + \text{h.c.} \, ,
\end{align}
where $\boldsymbol{h_{u,d,e}}$ are the usual $3\times 3$ Yukawa
matrices in flavor space (we will denote matrices in boldface).  In
turn, the VL fermions combined with $H_1$ induce
% -----------------------
% {\bf [In the Lagrangian as it is written in Avelino's
%   notes the different chiral components couple to the Higgs with
%   different strengths, resulting in four terms. Reshuffling these
%   terms I get two scalar ($\bar \psi\psi$) plus two pseudoscalar
%   ($\bar \psi\gamma_5\psi$) couplings, which in my opinion is not
%   consistent with these states being VL. I've changed the Lagrangian
%   accordingly (do you agree?)]}:
% -----------------------
\begin{align}
  \label{eq:Yukawa-Z4-inv}
  -\mathcal L_{Y}^{\text{VL}} &=
  \overline{Q}_L\cdot \boldsymbol{y_U}\cdot U_R\,\widetilde{H}_1
  + \overline{Q}_R\cdot \boldsymbol{\widetilde{y}_U}\cdot U_L\,\widetilde{H}_1
  \nonumber\\
  &+ \overline{Q}_L \cdot \boldsymbol{y_D} \cdot D_R\,H_1
  + \overline{Q}_R \cdot \boldsymbol{\widetilde y_D} \cdot D_L\,H_1
  % + \widetilde y_U \, \overline{Q_R} \widetilde{H}_2 U_L
  % \nonumber\\
  % &+ \widetilde y_D \, \overline{Q_R} H_2 D_L
  + \text{h.c.} \, ,
\end{align}
where $\boldsymbol{y_{U,D}}$ and $\boldsymbol{\widetilde y_{U,D}}$ are
$\nvl\times \nvl$ matrices in the VL flavor space.  Explicit mass
terms are given in turn by:
\begin{align}
  \label{eq:VL-quark-mass-terms}
  -\mathcal L_m &= \overline{Q}_L\cdot \boldsymbol{\hat m_Q}\cdot Q_R 
  + \overline{U}_L \cdot\boldsymbol{\hat m_U}\cdot U_R \nonumber \\
  & + \overline{D}_L \cdot \boldsymbol{\hat m_D}\cdot D_R
  + \text{h.c.} \, ,
\end{align}
where $\boldsymbol{\hat m_Q}$, $\boldsymbol{\hat m_U}$ and
$\boldsymbol{\hat m_D}$ are $\nvl\times \nvl$ matrices in the VL
flavor space which can be chosen to be diagonal without loss of
generality.

Leading order (LO) VL fermion effects are controlled by the
$\mathbb{Z}_2$-invariant renormalizable interactions
in~(\ref{eq:Yukawa-Z4-inv-SM})-(\ref{eq:VL-quark-mass-terms}). Higher
order explicit $\mathbb{Z}_2$-breaking effects are determined by
non-renormalizable operators. For $d_R=\boldsymbol{3}, \boldsymbol{6},
\boldsymbol{15}$ those effects are determined by the dimension-six
operators~\cite{Kumar:2011tj} \footnote{Higher order effective
  operators involving gluons are possible writing, see
  ref. \cite{Kumar:2011tj} for further details.}:
\begin{equation}
  \label{eq:eff-op-dimension-five}
  \mathcal{O}_6^{(1)}=\frac{C_6^{(1)}}{\Lambda^2}\,X_{d_R}\,q\,q\,\overline{q}\, ,
  \quad
  \mathcal{O}_6^{(2)}=\frac{C_6^{(2)}}{\Lambda^2}\,X_{d_R}\,q\,\overline{q}\,\overline{q}\, ,
\end{equation}
where $X_{d_R}$ stands for the VL colored fermion in the $d_R$
representation and $q$ refers to SM quark $SU(2)_L$ doublets or
singlets, depending on the electroweak charges of $X_{d_R}$. For
$d_R=\boldsymbol{8}, \boldsymbol{10}$, instead, effective LO effects
are given by~\cite{Kumar:2011tj}:
\begin{equation}
  \label{eq:eff-op-dimension-six}
    \mathcal{O}_6^{(1)\prime}=\frac{C_6^{(1)\prime}}{\Lambda^2}\,
    X_{d_R}\,q\,q\,q\, ,
  \qquad
  \mathcal{O}_6^{(2)\prime}=\frac{C_6^{(2)\prime}}{\Lambda^2}
  X_{d_R}\,\overline{q}\,\overline{q}\,\overline{q}\, .
\end{equation}
These operators are essential as they induce VL fermion decays, and so
are responsible for the signatures one could expect at the LHC (see
sec. \ref{sec:pheno}). Note however that when writing the effective
operators in (\ref{eq:eff-op-dimension-five}) and
(\ref{eq:eff-op-dimension-six}) one is implicitly assuming that the UV
completion can indeed lead to such effective interactions, so to a
large extent such effective approach is at any rate model-dependent.
If the LO effective effects are instead determined by a different set
of higher order effective operators (beyond six), the resulting
picture will of course be different.

A further constraint of the new states that one has to bear in mind
has to do with their contributions to electroweak precision data. Such
contributions have been studied in ref. \cite{Ellis:2014dza}, where is
has been shown---for the triplet case---that consistency with data it
is always achievable. In the case of higher-order color
representations one does not expect these conclusions to change since
these contributions are color-blind.

At the scalar level, the presence of the $\mathbb{Z}_2$ symmetry
constraints the scalar potential to have the form:
\begin{align}
  \label{eq:scalar-pot}
  \mathcal V = & m_{11}^2 H_1^\dagger H_1 + m_{22}^2 H_2^\dagger H_2 
  + \frac{\lambda_1}{2} \left(H_1^\dagger H_1\right)^2 
  + \frac{\lambda_2}{2} \left(H_2^\dagger H_2\right)^2 
  \nonumber \\
  & + \lambda_3 \left(H_1^\dagger H_1\right)\left(H_2^\dagger H_2\right) 
  + \lambda_4 \left(H_1^\dagger H_2\right)\left(H_2^\dagger H_1\right) 
  \nonumber \\
  & + \left[ \frac{\lambda_5}{2} \left(H_1^\dagger H_2\right)^2 
    + \,\text{h.c.} \right] \, .
\end{align}
\subsection{The CP-even Higgs mass matrix}
\label{sec:relevant-couplings}
We start by parametrizing the Higgs doublets according to:
\begin{equation}
  \label{eq:higgs-doublets-param}
  H_{1,2}=
  \begin{pmatrix}
    H_{1,2}^+\\
    H_{1,2}^0
  \end{pmatrix}\, ,
\end{equation}
where the neutral components are given by:
\begin{equation}
  \label{eq:neutral-com}
  H_{1,2}^0=\frac{1}{\sqrt{2}}
  \left(\varphi_{1,2}^0 + i\sigma_{1,2}^0 + v_{1,2}\right)\, ,
\end{equation}
with $\langle H_{1,2}^0\rangle=v_{1,2}/\sqrt{2}$, and $v^2=v_1^2+v_2^2\simeq
246\,$~GeV.  From the interactions in eq.~\eqref{eq:scalar-pot}, the CP
even mass matrix in the basis $(\varphi_{1}^0,\varphi_{2}^0)$ can be
written as:
\begin{equation}
  \label{eq:CP-even-Higss-matrix}
  \boldsymbol{\mathcal{M}}_H^2=
     \begin{pmatrix}
    m_{H_{11}}^2 &
    m_{H_{12}}^2\\
    m_{H_{12}}^2 &
   m_{H_{22}}^2
  \end{pmatrix}\, ,
\end{equation}
with the different entries, assuming $\lambda_5$ to be real, given by
\begin{align}
  \label{eq:mh-entries}
  m_{H_{11}}^2&=
  m_{11}^2+
  \frac{1}{2}
  \left[
    3\lambda_1v_1^2 +v_2^2
    \left(
      \lambda_3 + \lambda_4 + \lambda_5
    \right)
  \right]\, ,
  \nonumber\\
  m_{H_{12}}^2&=v_1v_2(\lambda_3+\lambda_4+\lambda_5)\, ,
  \nonumber\\
  m_{H_{22}}^2&=
  m_{22}^2+
  \frac{1}{2}
  \left[
    3\lambda_2v_2^2 +v_1^2
    \left(
      \lambda_3 + \lambda_4 + \lambda_5
    \right)
  \right]
  \, .
\end{align}
Defining the mass eigenstate basis according to\footnote{In this
  notation $h$ corresponds to the lightest CP even state.}
\begin{align}
  \label{eq:mass-gauge-eigenstate-basis}
  \begin{pmatrix}
    h\\
    H
  \end{pmatrix}
  &=
  \begin{pmatrix}
    -s_\alpha & c_\alpha\\
    c_\alpha & s_\alpha
  \end{pmatrix}
    \begin{pmatrix}
    \varphi_1\\
    \varphi_2
  \end{pmatrix} \equiv\boldsymbol{R_S}  \begin{pmatrix}
    \varphi_1\\
    \varphi_2
  \end{pmatrix}\,,
\end{align}
where $s_\alpha\equiv\sin\alpha$ and $c_\alpha\equiv\cos\alpha$, the
diagonalization of the matrix in eq.~\eqref{eq:CP-even-Higss-matrix}
proceeds as follows
\begin{equation}
  \label{eq:mass-matrix0-scalars}
  \boldsymbol{R_S}\cdot \boldsymbol{\mathcal{M}}_H^2\cdot 
  \boldsymbol{R_S}^\dagger
  =\boldsymbol{\hat{\mathcal{M}}}_H^2\, ,
\end{equation}
where the hat refers here and henceforth to diagonal matrices. The
mixing angle reads
\begin{equation}
  \label{eq:mixing_scalars}
  \tan2\alpha=\frac{-2\,m_{H_{12}}^2}{m_{H_{22}}^2-m_{H_{11}}^2}\, ,
\end{equation}
while the mass eigenvalues are given by (assuming for definiteness
$m_{H_{11}}^2>m_{H_{22}}^2$):
\begin{equation}
  \label{eq:masses_scalars}
  m_{h\,(H)}^2=\frac{1}{2}
  \left(
    \Delta m^2_{+} \mp 
    \,\sqrt{\Delta m^4_{-} + 4\,m_{H_{12}}^4}
  \right)\, ,
\end{equation}
with $\Delta m_{\pm}^2$ defined according to
\begin{equation}
  \label{eq:DeltamSq}
  \Delta m_\pm^2= m_{H_{11}}^2\pm m_{H_{22}}^2\, .
\end{equation}
Notice that in the limit $m_{H_{11}}^2\gg m_{H_{22}}^2>m_{H_{12}}^2$
we get:
\begin{align} \label{eq:masses_scalars_simple}
m_{h}^2& \simeq m_{H_{22}}^2-\frac{m_{H_{12}}^2}{m_{H_{11}}^2},\\
m_{H}^2&\simeq m_{H_{11}}^2\,,
\end{align}
and small mixing angle $\alpha$, see eq.~\eqref{eq:mixing_scalars}, so
that $h\sim \varphi_2,$ and $H\sim \varphi_1$. Finally, the CP-odd
Higgs mass can be written as
\begin{equation}
  \label{eq:CP-odd-scalar-mass}
  m^2_{A_0}=m^2_{H}+\lambda_5\,v_2^2\, .
\end{equation}
We stress that in the following we will fix $m_h=125$ GeV and
$1.8<m_H<2.2$ TeV.

As usual, the CP odd Higgses mass matrix is diagonalized by a $2\times
2$ unitary matrix parametrized with $\tan\beta=v_2/v_1$. Of relevance
for the process we will consider in Sec. \ref{sec:pheno} are the
couplings for $WWh$, $WWH$, $ZZh$ and $ZZH$ \cite{Gunion:2002zf}:
\begin{align}
  \label{eq:gauge-gauge-S-couplings}
  g_{VVh}=\frac{2M_V^2}{v}\,s_{\alpha-\beta}\, , \quad 
  g_{VVH}=\frac{2M_V^2}{v}\,c_{\alpha-\beta}\, .
\end{align}

\subsection{VL quark mass matrices}
\label{sec:mass-matrices}
In the presence of $\nvl$ VL fermion generations, the $%(
2\nvl%+3)
\times
%(
2\nvl%+3)
$ mass matrix%for the up-type quark sector
, written in the
left-right bases, with these bases defined as
$(\psi_{L,R}^U)^T=(\widetilde{U}_{L,R},U_{L,R}%,u_{L,R}
)$, reads:
\begin{equation}
  \label{eq:up-type-quarks-mass-matrix}
  \boldsymbol{\mathcal{M}_U} = 
  \begin{pmatrix}
    \boldsymbol{\hat m_Q} & \boldsymbol{\overline m_U}% & \boldsymbol{0}
    \\
    \boldsymbol{\tilde{\overline{m}}_{U}}^\dagger & 
    \boldsymbol{\hat m_U}% & \boldsymbol{0}
    \\
    %\boldsymbol{0} & \boldsymbol{0} & \boldsymbol{m_u}
  \end{pmatrix}\, ,
\end{equation}
where the following notation has been used:
\begin{eqnarray}
  \label{notation-mass-matrices-1}
  \boldsymbol{\overline m_{U}}&=&\frac{v}{\sqrt{2}} 
  c_\beta\,\boldsymbol{y_U}\, ,
  \qquad
  \boldsymbol{\tilde{\overline m}_{U}}=\frac{v}{\sqrt{2}} 
  c_\beta\,\boldsymbol{\tilde y_U}\, ,
%   \nonumber\\
%   \label{notation-mass-matrices-2}
%   \boldsymbol{\Lambda_{Qu}}&=&\frac{v}{\sqrt{2}}\,
%   \left(
%     c_\beta \boldsymbol{\lambda_{Qu}^{(1)}} 
%     + 
%     s_\beta \boldsymbol{\lambda_{Qu}^{(2)}}
%   \right)\, ,
%   \nonumber\\
%   \label{notation-mass-matrices-3}
%   \boldsymbol{\Lambda_U}&=&\frac{v}{\sqrt{2}}\,
%     \left(
%       c_\beta\boldsymbol{\lambda^{(1)}_{Uq}} 
%     + 
%     s_\beta\boldsymbol{\lambda^{(2)}_{Uq}}
%   \right)\, ,
%   \nonumber\\%[3mm\phantom]
%   \label{notation-mass-matrices-4}
%   \boldsymbol{m_u}&=&\frac{v}{\sqrt{2}}\,s_\beta \boldsymbol{h_u}\, .
\end{eqnarray}
%and $\boldsymbol{m_u}$ is the SM up-type quark mass matrix.  
The
down-type %quark 
sector mass matrix, in the bases
$(\psi_{L,R}^D)^T=(\widetilde{D}_{L,R},D_{L,R}%,d_{L,R}
)$, follows the
same structure, namely
\begin{equation}
  \label{eq:down-type-quarks-mass-matrix}
  \boldsymbol{\mathcal{M}_D} = 
  \begin{pmatrix}
    \boldsymbol{\hat m_Q} & \boldsymbol{\overline{m}_D} %&    \boldsymbol{0}
    \\
    \boldsymbol{\tilde{\overline{m}}_D}^\dagger & \boldsymbol{\hat m_D} %&      \boldsymbol{0}
    \\
    %\boldsymbol{0} & \boldsymbol{0} &     \boldsymbol{m_d}
  \end{pmatrix}\, .
\end{equation}
The parameters $\boldsymbol{\overline{m}_D}$ and
$\boldsymbol{\tilde{\overline{m}}_D}$ are given by those in
(\ref{notation-mass-matrices-1}) by trading the subindex $U\to D$. 
% As in the up-type sector, the  $3\times 3$
% matrix in the lower right
% corner refers to the SM down-type quark mass matrix.

%% SM quarks and VL quark mixing is encoded in the submatrices:
%% \begin{equation}
%%   \label{eq:submatrices-mix}
%%   \boldsymbol{m_{RVL}^{(U,D)}}=
%%   \begin{pmatrix}
%%     \boldsymbol{\Lambda_{Q(u,d)}}\\
%%     \boldsymbol{\mu_{(U,D)}}
%%   \end{pmatrix}\, ,
%%   \quad
%%   \boldsymbol{m^{(U,D)}_{LVL}}^\dagger=
%%   \begin{pmatrix}
%%     \boldsymbol{\mu_Q}^\dagger & \boldsymbol{\Lambda_{(U,D)}}^\dagger
%%   \end{pmatrix}\ ,
%% \end{equation}
%% where the matrix $\boldsymbol{m^{(U,D)}_{RVL}}$
%% ($\boldsymbol{m^{(U,D)}_{LVL}}$) accounts for mixing between VL and SM
%% right-handed (left-handed) quarks.

Defining the mass eigenstate bases as
\begin{equation}
  \label{eq:mass-eigenstates}
  \Psi_L^{(U,D)}=\boldsymbol{%V
                            R_L^{(U,D)}}\,\psi_L^{(U,D)}\, ,
  \quad
  \Psi_R^{(U,D)}=\boldsymbol{%V
                            R_R^{(U,D)}}\,\psi_R^{(U,D)}\, ,
\end{equation}
both matrices can therefore be diagonalized through biunitary
transformations:
\begin{equation}
  \label{eq:diagonalization}
  \boldsymbol{%V
              R_L^{(U,D)}}\cdot\boldsymbol{\mathcal{M}^{(U,L)}}\cdot
  \boldsymbol{%V
              R_R^{(U,D)\dagger}}
  =\boldsymbol{\hat{\mathcal{M}}^{(U,L)}}\, .
\end{equation}

\subsection{Relevant Higgs couplings}
\label{sec:relevant-couplings2}
Recasting the interactions in eq.~\eqref{eq:Yukawa-Z4-inv} in the mass
eigenstates bases for both, the VL fermions and $H_1$, one gets for the
LR couplings
\begin{align}
  \label{eq:LR-Lag-mass-eig}
  \mathcal{L}_{LR}&= \sum_{a,b=1}^{2\nvl}
    \overline{\Psi}_{L_a}^U\,O^{ULR}_{ab}\,\kappa_A\,\Psi_{R_b}^U\,S_A
    \nonumber\\
    &+
    \sum_{a,b=1}^{2\nvl}
    \overline{\Psi}_{L_a}^D\,O^{DLR}_{ab}\,\kappa_A\,\Psi_{R_b}^D\,S_A
    + \mbox{h.c.}\, ,
\end{align}
where for $A=1$, $\kappa_1=-s_\alpha$ and $S_1=h$, while for $A=2$,
$\kappa_2=c_\alpha$ and $S_2=H$. The couplings for the up- and
down-type sectors are given by
\begin{align}
  \label{eq:LR-couplings}
  O^{ULR}_{ab}&=R^U_{L_{ac}}\,Y_{U_{cd}}\,R^{U*}_{R_{bd}}\, ,
  \nonumber\\
  O^{DLR}_{ab}&=R^D_{L_{ac}}\,Y_{D_{cd}}\,R^{D*}_{R_{bd}}\, .
\end{align}
Here $\boldsymbol{Y_{X_{cd}}}$ are the elements of the $2 \nvl \times
2 \nvl$ matrix
\begin{equation}
\boldsymbol{Y_X}=
\begin{pmatrix}
  \boldsymbol{0}_{\nvl} & \boldsymbol{y_X} \\
  \boldsymbol{\tilde{y}^\dagger_X}& \boldsymbol{0}_{\nvl}
\end{pmatrix}\, ,
\end{equation}
with $X=U,D$ and $\boldsymbol{0}_{\nvl}$ a $2\nvl \times 2\nvl$ matrix
with vanishing elements. Summation over repeated indices is assumed in
eq. \eqref{eq:LR-couplings}.

A simple case of interest for our phenomenological analysis is that
where $\nvl=1$. In that case, the couplings $y_X$ and
$\widetilde{y}_X$ can be taken to be real without loss of
generality. Thus, the matrices $\boldsymbol{R^{(U,D)}_{L,R}}$ can be
parameterized according to
\begin{equation}
  \label{eq:Rot-R-U-D}
  \boldsymbol{R^{(X)}_{L,R}}=
  \begin{pmatrix}
    \cos\theta^{X}_{L,R} & \sin\theta^X_{L,R}\\
    -\sin\theta^X_{L,R} & \cos\theta^X_{L,R}
  \end{pmatrix}
  \qquad (X=U,D)\, ,
\end{equation}
with the corresponding mixing angles given by
\begin{align}
  \label{eq:mixing-angle-one-generation}
  \tan2\theta_L^X&=-2
  \frac{
    m_Q\,\widetilde{\overline m}_X
      +
      m_X\,\overline{m}_X
    }{m_X^2 - m_Q^2 - \overline{m}_X^2 + \widetilde{\overline{m}}_X^2}\, ,
  \nonumber\\
  \tan2\theta_R^X&=-2
  \frac{
    m_Q\,\overline{m}_X
      +
      m_X\,\widetilde{\overline m}_X
    }{m_X^2 - m_Q^2 - \widetilde{\overline{m}}_X^2 + \overline{m}_X^2}\, .
\end{align}
%\draftnote {I propose to change this blue to:\\}
% Furthermore, being orthogonal $2\times 2$ matrices (see matrix
% $\boldsymbol{R}$ in eq. \ref{eq:mass-gauge-eigenstate-basis}), they
% can be parameterized by a single angle $\theta_X$ (as in
% eq. \ref{eq:mixing_scalars}), which for $\hat m_Q>\hat m_X$ reads:
% \begin{align}
%   \label{eq:mixing-angle-one-generation}
%   \tan2\theta_X&=\frac{2\overline{m}_X}{\hat m_Q-\hat m_X}\, ,
% \end{align} 
% with $X=U,D$.
With the aid of eq. (\ref{eq:Rot-R-U-D}), the interactions in
eq.~\eqref{eq:LR-Lag-mass-eig} written in the mass eigenstate basis
are given by
\begin{align}
  \label{eq:Lag-up-type-nVLEq1}
  \mathcal{L}_U&=
  \frac{\kappa_A}{\sqrt{2}}\,
  \left(
    Y_{11}^U\,\overline{\widetilde{U}'}_L\,\widetilde{U}_R'
    +
    Y_{12}^U\,\overline{U}'_L\widetilde{U}'_R
  \right.
  \nonumber\\
  &\left.
    + Y_{21}^U\,\overline{\widetilde{U}'}_LU'_R
    + Y_{22}^U\,\overline{U}_L'\,U_R'
  \right)\,S_A + \mbox{h.c.}\, ,
\end{align}
where the different couplings read:
\begin{align}
  \label{eq:y1}
  Y_{11}^U&=y_U \cos\theta_L^U \sin\theta_R^U + \widetilde{y}_U \cos\theta_R^U
  \sin\theta_L^U\, ,
  \\
  \label{eq:Y12}
  Y_{12}^U&=
  y_U\cos\theta_L^U\cos\theta_R^U - \widetilde{y}_U\sin\theta_L^U\sin\theta_R^U\, ,
  \\
  \label{eq:Y21}
  Y_{21}^U&=-y_U\sin\theta_L^U\sin\theta_R^U + \widetilde{y}_U\cos\theta_L^U\cos\theta_R^U
  \, ,
  \\
  \label{eq:y2}
  Y_{22}^U&=-\left(y_U \sin\theta_L^U \cos\theta_R^U 
    + \widetilde{y}_U \cos\theta_L^U\sin\theta_R^U\right)\, .
\end{align}
Those in the down-type sector have the form
\begin{align}
  \label{eq:Lag-down-type-nVLEq1}
  \mathcal{L}_D&=
  \frac{\kappa_A}{\sqrt{2}}
    \left(
    Y_{11}^D\,\overline{\widetilde{D}'}_L\,\widetilde{D}_R'
    +
    Y_{12}^D\,\overline{D}'_L\widetilde{D}'_R
  \right.
  \nonumber\\
  &\left.
    + Y_{21}^D\,\overline{\widetilde{D}'}_LD'_R
    + Y_{22}^D\,\overline{D}_L'\,D_R'
  \right)\,S_A + \mbox{h.c.}\, ,
\end{align}
with the down-type sector couplings given as in
(\ref{eq:y1})-(\ref{eq:y2}) trading $U\to D$.  The primes refer to
the fields written in the mass basis.

As we have already pointed out, the symmetry transformations in
eq.~(\ref{eq:transformations}) allow the chiral (SM) and VL sectors to
be decoupled in such a way that VL couplings can not sizeably affect
the SM Higgs single production cross-section. This can be seen in
tab. \ref{tab:couplings}, where we have listed the couplings of both
sectors according to the interactions in (\ref{eq:Lag-up-type-nVLEq1})
and (\ref{eq:Lag-down-type-nVLEq1}). In the limit $\sin\alpha\to 0$
any such contribution will vanish, while those related with the heavy
CP-even scalar $H$ will be enhanced. This is, in our opinion, an
interesting feature of our setup: the condition of large contributions
to the $H$ single production cross-section assures negligible (or even
vanishing) contributions to the SM $h$ single production.\\

\begin{table}
  \centering
  \setlength{\tabcolsep}{0.2cm}
  \renewcommand{\arraystretch}{1.3}
  \begin{tabular}{|c|c|c|}
    \hline
    \cline{1-3}
    \multicolumn{3}{|c|}{\textbf{Standard Model}}
    \\\hline
    \textbf{Couplings} & $\boldsymbol{u}$ \textbf{sector} $(f=u)$ 
    & $\boldsymbol{d}$ \textbf{sector} $(f=d)$
    \\\hline
    $g_{f_if_ih}$ & $(m_{u_i}/v)(c_\alpha/s_\beta)$ 
    & $(m_{d_i}/v)(\,c_\alpha/s_\beta)$
    \\\hline
    $g_{f_if_iH}$ & $(m_{u_i}/v)(\,s_\alpha/s_\beta)$ 
    & $(m_{d_i}/v)(\,s_\alpha/s_\beta)$
    \\\hline\hline
    \cline{1-3}
    \multicolumn{3}{|c|}{\textbf{Vector-like}}
    \\\hline
    \textbf{Couplings} & $\boldsymbol{U}$ \textbf{sector} $(F=U)$ 
    & $\boldsymbol{D}$ \textbf{sector} $(F=D)$
    \\\hline
    $g_{F_iF_jh}$ & $Y_{ij}^U\,s_\alpha$ 
    & $Y_{ij}^D\,s_\alpha$ 
    \\\hline
    $g_{F_iF_jH}$ & $Y_{ij}^U\,c_\alpha$ 
    & $Y_{ij}^D\,c_\alpha$
    \\\hline
  \end{tabular}
  \caption{Yukawa couplings for SM and VL up- and down-type quarks. 
    Note that enhanced $g_{F_iF_jH}$ couplings 
    ($c_\alpha\to 1$) guarantee negligible $g_{F_iF_jh}$ parameters.}
      \label{tab:couplings}
    \end{table}
    
\section{Phenomenological analysis}
\label{sec:pheno}
Higgs properties derived from production and decay mode analyses at
LHC have placed stringent bounds on 2HDMs
\cite{Carmi:2012in,Eberhardt:2013uba,Barroso:2013zxa}. Although
consistency with data still allows for certain freedom, favored
regions in parameter space are those corresponding to the decoupling
limit~\cite{Carmi:2012in}, in which apart from $h$ (whose mass is
fixed to $\sim$125~GeV~\cite{Aad:2012tfa,Chatrchyan:2012xdj}) the
remaining part of the scalar mass spectrum is heavy
\cite{Gunion:2002zf}. In terms of the scalar sector mixing angles,
this translates into small $\cos(\alpha-\beta)$, with the possible
values for $\tan\beta$ depending on the model itself
\cite{Barroso:2013zxa,Ferreira:2014sld}. For the type-I 2HDM, which
corresponds in our case to the SM sector, values of
$\sin(\alpha-\beta)$ close to 1 do not necessarily demand large values
of $\tan\beta$, as it turns out to be e.g. in the type-II 2HDM
\cite{Ferreira:2014sld}.

The heavier CP-even state $H$ can be produced solely through SM
interactions, as can be noted from tab. \ref{tab:couplings}. However,
the corresponding cross-section in that case is expected in general to
be small, as can be seen by going to the decoupling limit.  Since in
that case $\alpha=\beta+\pi/2$, $g_{tth}$ matches the SM coupling. The
coupling $g_{ttH}$, instead, becomes $(m_t/v)\,\cot\beta$, which even
for moderate values of $\tan\beta$ implies a suppressed production
cross-section. Thus, sizeable production of $H$ is only possible
through the VL couplings. As can be seen in tab.~\ref{tab:couplings},
both $g_{FFh}$ and $g_{FFH}$ in that case are not sensitive to values
of $\tan\beta$, suppressed (enhanced) production of $h$ ($H$) can be
achieved solely through small (large) values of $s_\alpha$
($c_\alpha$). Thus, one can consistently get enhanced $H$ production
without considerably affecting $h$ production.
% Something that one
% should bear in mind is that although $g_{FFH}$ does not depend upon
% $\tan\beta$, the condition of being close to the decoupling limit
% implies that enhanced $H$ production requires values of $\tan\beta$
% larger than 1 (see discussion below).

The gluon fusion $H$ production cross-section strongly depends on the
VL fermion mass spectrum. The dependence enters in two ways, namely.
The loop function combined with the fact that in this case the Yukawa
couplings are not directly related with the VL fermion masses, induce
a decoupling behavior which for heavy VL mass spectra strongly
suppresses the cross-section. Secondly, depending on the VL and scalar
mass spectra, the new states can sizeably contribute to the running of
$\alpha_{\rm s}$, largely changing its value (for more details see
sec. \ref{sec:pheno-analy}).

In summary, as what regards $H$ production, consistency with data
requires being close to the decoupling limit. This condition implies
suppressed $g_{ffH}$ couplings which then demands $H$ production
through VL couplings. The contributions of these couplings to the
gluon fusion $h$ production cross-section are small (or can even
vanish), so it is possible to achieve a consistent picture of
``large'' $H$ production.

CP-odd scalar production proceeds in the same way, controlled by the
same set of parameters. The only difference resides in the mass
difference between the heavy CP-even and the CP-odd scalar, which due
to the constraints implied by the scalar potential are small: taking
the non-perturbative limit value $\lambda_5=4\pi$, one gets the bound
(see eq. (\ref{eq:CP-odd-scalar-mass})):
\begin{equation}
  \label{eq:cp-odd-mass-limit}
  m_A\lesssim 2.3\;\mbox{TeV}\, ,
\end{equation}
a value in agreement with electroweak precision data.

% VL quarks necessarily contribute to the inclusive cross-section of the
% heavy Higgs ($m_H\subset [1.8,2]\,$~TeV). The exact value depends
% mainly on two parameters, the $Y_{QQH}$ Yukawa coupling and the VL
% quarks masses $m_Q$ (see below for further details). In contrast to
% chiral quarks, these parameters are not related and so can be treated
% independently. Of particular relevance is $m_Q$, since the cross
% section induced by these states exhibits a decoupling behavior which
% one would n\"aively expect to render the corresponding cross-section
% small. This is in indeed the case for $m_Q\gtrsim1.2\,$~TeV, but is
% not in the ``low'' VL quark mass ``regime'' where the interplay
% between $Y_{QQH}$ and $m_Q$ yields production cross-sections well in
% the required range, ${\cal O}(\sigma^{pp}_{H_2})\sim 10\,$~fb
% \cite{Allanach:2015hba}.

\subsection{Production cross-section and VL fermion mass limits}
\label{sec:pheno-analy}
In the case $n_{VL}=1$ there are two different contributions to
$\sigma(pp\to H)$ for both, the up- and down-type VL sectors. The
contributions are determined by the first and fourth terms in
eqs. (\ref{eq:Lag-up-type-nVLEq1}) and
(\ref{eq:Lag-down-type-nVLEq1}). The gluon fusion cross-section then
has the form
\begin{align}
  \label{eq:x-sec}
  \hat \sigma(gg\to H)&=
  \frac{\kappa\,\alpha_{\rm s}^2}{64\,\pi} \cos^2 \alpha
  \left|T_R\,\sum_{\substack{X=U,D\\i=1,2}}
    \frac{Y_{ii}^X}{\sqrt{2}}\frac{A(m_H^2/m_{X_i}^2)}{m_{X_i}}
  \right|^2
  \nonumber\\
  &\times\tau_0 \, \delta\left(\tau - \tau_0\right)\ .
\end{align}
Here we have included a $\kappa$ factor to account for NLO corrections
and defined $\tau = s/S$ and $\tau_0=m_H^2/S$, where $s$ and $S$ are
the parton-parton and proton-proton center of mass energies,
respectively. In our numerical analysis we will fix $\sqrt{S} = 8$
TeV. The ``effective'' couplings $Y^{U,D}_{ii}$ are given by
eqs. (\ref{eq:y1}) and (\ref{eq:y2}) and encode the dependence of the
cross-section upon the VL Yukawa couplings and VL fermion mixing. The
loop function $A(m_H^2/m_{X_i}^2)$ reads
\begin{equation}
  \label{eq:loop-fun}
  A(m_H^2/m_{X_i}^2)=\int_0^1\,dy\,\int_0^{1-y}\,dz\,
  \frac{1-4yz}{1-(m_H^2/m_{X_i}^2)yz}\, ,
\end{equation}
and match after integration the standard one-loop functions for Higgs
production via gluon fusion (see e.g. ref.~\cite{Djouadi:2005gi}).

The physical cross-section at the LHC requires integration over the
parton distribution functions $g(x)$:
%\begin{equation}
%  \label{eq:parton-distribution-xsec}
%  \sigma(pp\to H)=\sigma(gg\to H)\,\tau_0\,
%  \int_{y_0}^{-y_0}\,dy\,g(y_1)
%  g(y_2)\, ,
%\end{equation}
%where the following conventions have been adopted:
%$y_1=\sqrt{\tau_0\,e^y}$, $y_2=\sqrt{\tau_0\,e^{-y}}$,
%$y_0=\log\sqrt{\tau_0}$ and $\tau_0=m_H^2/s$.
\begin{equation}
  \label{eq:parton-distribution-xsec}
  \sigma(pp\to H)=\int_{\tau_0}^1 dx_1 \int_{\tau_0/x_1}^1 dx_2 \, g(x_1) \, g(x_2) \, \hat \sigma(gg\to H) \, .
\end{equation}
%where $\tau_0=m_H^2/s$.

Some words are in order regarding the group theory factors $T_R$. The
gluon-VL-VL coupling structure is determined by the $SU(3)_c$
generators $t^a_R$, which in turn depend upon the VL irreducible
representation $R$, assumed to be of rank $(\lambda_1,\lambda_2)$. The
amplitude for the gluon fusion process, therefore, involves
$\mbox{Tr}(t^a_R\,t^b_R)$, whose value is given by the trace
normalization condition:
\begin{equation}
  \label{eq:trace}
  \mbox{Tr}(t^a_R\,t^b_R)=\frac{C_R\,d_R}{d_A}\,\delta^{ab}=T_R\,\delta^{ab}\, ,
\end{equation}
where $d_R$ and $d_A$ refer to the dimensions of the representation
$R$ and the adjoint ($A=\boldsymbol{8}$), and $C_R$ is the constant
that defines the quadratic Casimir, namely
\begin{align}
  \label{eq:casimir}
  d_R&=\frac{1}{2}(\lambda_1 + 1)(\lambda_2 +1)(\lambda_1 + \lambda_2
  +2)\, ,
  \nonumber\\
  C_R&=\frac{1}{3}(\lambda_1^2 + \lambda_2^2 + 
  \lambda_1\lambda_2 + 3\lambda_1 + 3\lambda_2)\, .
\end{align}
Bearing in mind that the adjoint is rank (1,1), $T_R$ is entirely
determined by the rank of the corresponding representation, namely:
$\boldsymbol{3}=(1,0)$, $\boldsymbol{6}=(2,0)$,
$\boldsymbol{10}=(3,0)$ and $\boldsymbol{15}=(2,1)$. Values for $d_R$,
$C_R$ and $T_R$ for the lower-dimensional $SU(3)$ representations are
given in tab. \ref{tab:casimir}.

We discuss now the evolution of $\alpha_{\rm s}$ under the
Renormalization Group Equations (RGEs).  In the absence of the new
colored states (SM alone) we find $\alpha_{\rm s}(M_Z)/\alpha_{\rm
  s}(m_H) = \left\{1.40, 1.42, 1.43 \right\}$ for $m_H = \left\{ 1.8,
2.0, 2.2 \right\}\,$~TeV (the values used in our numerical
treatment). The VL fermions, in particular those belonging to higher
order color representations, can substantially change those values
through their non-negligible positive contributions to the RGE
running. Whether this is the case depends on the corresponding VL mass
spectrum. For spectra heavier than $m_H$ there is no contribution,
thus the values previously quoted are the ones to be used. For spectra
with at least one VL state with mass below $m_H$, the $\alpha_{\rm s}$
RGE running should be accounted for, as it may have a non-negligible
numerical impact on the resulting $H$ production cross-section. The
$\alpha_{\rm s}$ RGE reads:
\begin{equation}
  \label{eq:beta}
  \mu \frac{d\alpha_{\rm s}}{d\mu} = \alpha_{\rm s}\,\sum_i \,\beta_i \,
  \left(\frac{\alpha_{\rm s}}{\pi}\right)^i\, ,
\end{equation}
with the one- and two-loop $\beta$ functions given by
\cite{Caswell:1974gg,Jones:1974mm}
\begin{align}
  \label{eq:beta-1}
  \beta_1&=-\frac{11}{6}\,C_{\rm A} 
  +\frac{2}{3}\,\sum_{\rm R}\,n_{\rm R}\,T_{\rm R}\, ,\\
   \beta_2&=-\frac{17}{12}\,C_{\rm A} ^2 
   +\frac{1}{6}\,\sum_{\rm R} \,n_{\rm R}\,T_{\rm R}\,(5\,C_{\rm A}+3\,C_{\rm R})\, ,
\end{align}
where $n_{\rm R}$ is the number of quark flavors in the representation
$R$. Clearly, the higher the rank of the representation (large $T_R$ and $C_R$),
the larger the contribution of the VL states to $\alpha_{\rm
  s}$. Indeed, it can be noted that for higher rank representations a
Landau pole will be reached rapidly, implying in those cases the need
for further color states with order-TeV masses, so to assure a good UV
behavior.

Having accounted for all the relevant effects, it becomes clear that
the gluon fusion Higgs production cross-section for different representations
differ solely by the group theory factor and the value of
$\alpha_{\rm s}$ at $m_H$. Thus, as soon as its value is determined
for a particular representation, values for the others can be
straightforwardly derived by rescaling by the appropriate
factors. Relative to the fundamental representation,
$F=\boldsymbol{3}$, these factors are %DR changes in blue.
\begin{equation}
  \label{eq:enhancement-fac-H}
  \varepsilon_{gg}^H\equiv\frac{\sigma_\text{R}(gg\rightarrow H)}
  {\sigma_\text{F}(gg\rightarrow H)}=
  \frac{T_R^2}{T_F^2}\,\frac{\alpha_{\text{s},R}^2}{\alpha_{\text{s},F}^2}\, .
\end{equation}
    
Tab. \ref{tab:casimir} shows the group theory enhancements for the
irreducible representations of interest, from which it can be seen
that large cross-sections are expected for higher-rank
representations. This is however subtle, since large cross-sections
demand not too heavy VL states. And is for higher rank representations
for which one could expect the most stringent bounds on their masses.
This statement is, however, to a large extent, model-dependent. Bounds
are derived assuming certain VL fermion decay modes, which in the
absence of chiral-VL mixing entirely depend upon the effective
operator assumed, as we now discuss.

\begin{table}
  \centering
  \setlength{\tabcolsep}{0.2cm}
  \renewcommand{\arraystretch}{1.3}
  \begin{tabular}{|c|c|c|c|c|c|}
    \hline
    \cline{1-6}
    \multicolumn{6}{|c|}{$SU(3)_{\rm c}$ \textbf{representations}}
    \\\hline\cline{1-6}
    $d_{\rm R}$ & $\boldsymbol{3}$ & $\boldsymbol{6}$&$\boldsymbol{8}$& 
    $\boldsymbol{10}$&$\boldsymbol{15}$ \\\hline
    $C_{\rm R}$ & $4/3$ & $10/3$ & $3$ & $6$&$16/3$ \\\hline
    $T_{\rm R}$ & $1/2$ & $5/2$ & $3$ & $15/2$&$10$ \\\hline
    $\varepsilon_{gg}^H\,\times \,\alpha^2_{\rm S,F}/\alpha^2_{\rm S,%X
      R}$
    & $1$ & $25$ & $36$ & $225$&$400$ \\\hline
    % $\varepsilon_{qq}^\text{VL}$
    % & $1$ & $5$ & $6$ & $15$&$20$ \\\hline
    $\varepsilon_{gg}^\text{(%X
      R,F)}$  & 
    $1$ & $25/2$ & $27/2$ & $135/2$&$80$ \\\hline
  \end{tabular}
  \caption{Dimension $d_{\rm R}$, quadratic Casimir
    operators coefficient ($C_R$) and trace factor ($T_R$)
    for lower-dimensional color representations. The
    coefficient $\varepsilon_{gg}^H$ refers to the enhancement
    of the heavy Higgs gluon fusion cross-section for
    representations $R= \boldsymbol{6}, \boldsymbol{8},
    \boldsymbol{10}, \boldsymbol{15}$, relative to the fundamental
    representation ($F=\boldsymbol{3}$). The other coefficient 
    $\varepsilon_{gg}^\text{(F,X)}$ refers instead to the enhancement 
    of the VL pair production cross-section $\sigma(gg \to X\bar X)$.
    Note that, in the Higgs gluon fusion enhancement coefficient,
    $\alpha_{\rm s}$ has been evaluated at $m_H$. See 
    text for further details.}
  \label{tab:casimir}
\end{table}

VL fermions pair-production is mainly driven by $gg\to X\bar X$. At
leading order in $m_X^2/S$, the production cross-sections for 
representations $R_a$ and $R_b$ differ by~\cite{Ilisie:2012cc}:
\begin{equation}
  \label{eq:ratio-pair-prod}
  \varepsilon_{gg}^{(R_a,R_b)}\equiv
  \frac{\sigma(gg\to X_a\bar X_a)}{\sigma(gg\to X_b\bar X_b)}
  \simeq\frac{C_{R_a}^2}{C_{R_b}^2}\,\frac{d_{R_a}}{d_{R_b}}\, .
\end{equation}
Such values are shown in tab. \ref{tab:casimir} relative to the
fundamental representation.  With the aid of these rescaling factors,
bounds on the masses of different representations can be indirectly
estimated from experimental bounds on the mass of a given one.  Such
an approach assumes the VL fermions to be short-lived, with lifetimes
below 10~ns. For lifetimes above this value (and below $\sim 100$~s,
as required by cosmological and astrophysical constraints
\cite{Mack:2007xj}), the VL fermions would be stable or metastable
depending on whether they decay outside or inside the active detector
volume \cite{Aad:2015qfa}. In that case, arguably, bounds on the
different VL fermions could be fixed by using current bounds on
charged heavy long-lived particles, for which current bounds exclude
masses below $\sim 1\,$~TeV \cite{Aad:2015qfa}.

VL fermion lifetimes are determined by the effective operator
responsible for its decay (see sec. \ref{sec:vl-quarks}). Assuming
$\mathcal{O}(C_6, C_6^\prime)\sim 1$ and taking $m_X=1.5\,$~TeV, we
have found that short-lived VL fermions are obtained for cutoff scales
obeying $\Lambda\lesssim 10^7\,$~GeV, stability at collider scales is
instead obtained for $10^{7}\,\mbox{GeV}\lesssim\Lambda\lesssim
10^{10}\,\mbox{GeV}$ (where the upper bound assures decay lifetimes
below 100 s).

In the short-lived case, mass limits for the different representations
can be derived by using current bounds on gluino masses in models with
R-parity violation \cite{Aad:2015lea}. These limits, derived from
searches for six jets stemming from R-parity-violating gluino decays,
have excluded gluino masses below $\sim900\,$~GeV.  Since the gluino
is a VL octet, these bounds combined with appropriate rescaling
factors can---in principle---be used to derive lower limits on the
remaining VL quark representations, provided the VL decay modes yield
a six jet topology. This is indeed the case for decays induced by the
effective operators in~(\ref{eq:eff-op-dimension-five}) and
(\ref{eq:eff-op-dimension-six}). Thus, in that case for
\begin{alignat}{2}
  \label{eq:mass-limits}
  \varepsilon_{gg}^{(F,A)}&=\frac{2}{25}\, ,\qquad&
  \varepsilon_{gg}^{(6,A)}&=\frac{25}{27}\, ,
  \nonumber\\
  \varepsilon_{gg}^{(10,A)}&=\frac{27}{5}\, ,\qquad&
  \varepsilon_{gg}^{(15,A)}&=\frac{32}{5}\, , %
\end{alignat}
we find no competitive bound for $m_{X_3}$ (so its lower value is then
fixed to 500~GeV~\cite{Okada:2012gy}), for $m_{X_6}\gtrsim 833\,$~GeV
and for $m_{X_{10},X_{15}}\gtrsim 4000\,$~GeV. The latter, being at
the LHC kinematical threshold, is therefore expected to be somewhat
degraded.

The octet having a weaker mass bound and an expected large
cross-section, it is probably the most suitable VL fermion for
addressing the ATLAS diboson excess (in the case of short-lived VL
fermions). Thus, most of our results in sec.~\ref{sec:numerics} will
specialize to this case.

Finally, before closing this section it is worth pointing out that the
above limits imply a depletion of the gluon fusion cross-section for
the different representations, apart from the triplet for which less
stringent bounds apply. This is to be compared with the case where the
states rather than being VL are chiral, since constraints on a fourth
chiral quark generation are less restrictive, $m_\text{chiral}\gtrsim
600\,$~GeV \cite{Chatrchyan:2012fp}\footnote{Within our setup, a fourth
  chiral generation coupling only to $H_1$ is absolutely viable, since
  its contribution to the SM Higgs cross-section is negligible or even
  vanishing.}, and the cross-section does not exhibit a decoupling
behavior. Note, however, that in that case addressing the diboson
anomaly is not possible: the large Yukawa couplings required to
generate experimentally consistent chiral masses necessarily lead to
a heavy Higgs total decay width above $\sim200$ GeV(see next section).
\subsection{Partial decay widths}
\label{sec:decay-widths}
The dominant $H$ decay modes are: $H\to V\,V$ ($V=W,Z$), $H\to
X_i\bar{X}_j$ (see ecs. (\ref{eq:Lag-up-type-nVLEq1}) and
(\ref{eq:Lag-down-type-nVLEq1})).  The partial decay widths for these
processes can be written as:
\begin{align}
  \label{eq:partial-decay-widths-gauge}
  \Gamma(H\to VV)&=\frac{\delta_VG_F}{16\sqrt{2}\pi}m_H^3c^2_{\alpha-\beta}
  \,G(1,r_{VH}^2,r_{VH}^4)\, ,
  \\
  \label{eq:partial-decay-widths-VL-non-mixed}
  \Gamma(H\to X_i\,\bar X_i)&=\frac{d_R}{8\pi}\frac{\left|Y_{ii}^{X}\right|^2}{2}\,m_H
  \,\lambda^{3/2}(1,r_{X_iH}^2,r_{X_iH}^2)\, ,
  %\left(1 - 4\frac{m_{Q_i}^2}{m_H^2}\right)^{3/2}\, ,
  \\
  \label{eq:partial-decay-widths-VL-mixed}
  \Gamma(H\to X_i\,\bar X_j)&=\frac{d_R}{8\pi}\frac{\left|Y_{ij}^{X}\right|^2}{2}\,m_H\,
  F(1,r_{X_1H}^2,r_{X_2H}^2)\, .
\end{align}
Here $\delta_V=2$ for $V=W$ and $\delta_V=1$ for $V=Z$, $i=1,2$ refer
to the states in the up and down sector, $Y_{ij}^X$ are the
off-diagonal couplings given in (\ref{eq:Y12}) and (\ref{eq:Y21}),
$r_{VH}=m_V/m_H$, $r_{X_iH}=m_{X_i}/m_H$ and the kinematic functions
read
\begin{align}
  \label{eq:kinematic-fun}
  G&=(1 -4r_{VH}^2 + 12r_{VH}^4)\lambda^{1/2}(1,r_{VH}^2,r_{VH}^2)\, ,
  \nonumber\\
  F&=
  \left[1 - (r_{x_1 H} + r_{x_2 H})^2\right]\,
  \lambda^{1/2}(1,r_{x_1 H}^2,r_{x_2 H}^2)\, ,
\end{align}
with $\lambda(1,a,b)=1+(a-b)^2-2(a+b)$.

Depending on the relative size of $\cos(\alpha-\beta)$ and the
``effective'' couplings $Y_{ij}^{U,D}$ ($i=1,2$), the total decay
width is controlled either by gauge boson modes or fermion decays.
For higher representations, fermion decay dominance is more pronounced
due to the largest number of color degrees of freedom. However, in
general, for ``effective'' couplings $Y_{ij}^{U,D}$ smaller than one
the gauge boson modes dominate, unless $\cos(\alpha-\beta)\lesssim
0.01$. On the contrary, for large ``effective'' couplings, fermion
modes can determine the total decay width, with the
corresponding value typically being well above $\sim100$ GeV.
\subsection{Numerical results}
\label{sec:numerics}
No matter the color representation, the heavy Higgs gluon fusion
cross-section depends on the VL mixing as well as on the VL fermion
mass spectrum (in both up- and down-type sectors). Their values are to
a large extent correlated since they both depend upon the same set of
Lagrangian parameters, and although they depend differently there is
no room for variations on the mixing giving a mass spectrum. Thus,
rather than treating mixing and spectrum independently, in our
analysis we used the ``fundamental'' couplings, assuming common values
for both sectors. Such an assumption certainly simplifies the
numerical treatment, while capturing the main features of the
parameter space dependence. Our results are therefore derived for
fixed $m_H=1.8\,$~TeV and $\sqrt{S}=8\,$~TeV and are based on random
scans of the following parameter space regions:

\begin{alignat}{2}
  \label{eq:regions}
  m_{Q,X} &\subset [500,2500]\,\mbox{GeV}\, ,\qquad
  &&y,\widetilde{y}\subset [10^{-1},\sqrt{4\pi}]\, ,
  \nonumber\\
  \tan\beta&\subset [0.3,10]\, ,
  \qquad
  &&\sin(\alpha-\beta)\subset [0.9,1]\, .
\end{alignat}
\begin{figure}
  \centering
  \includegraphics[scale=0.3]{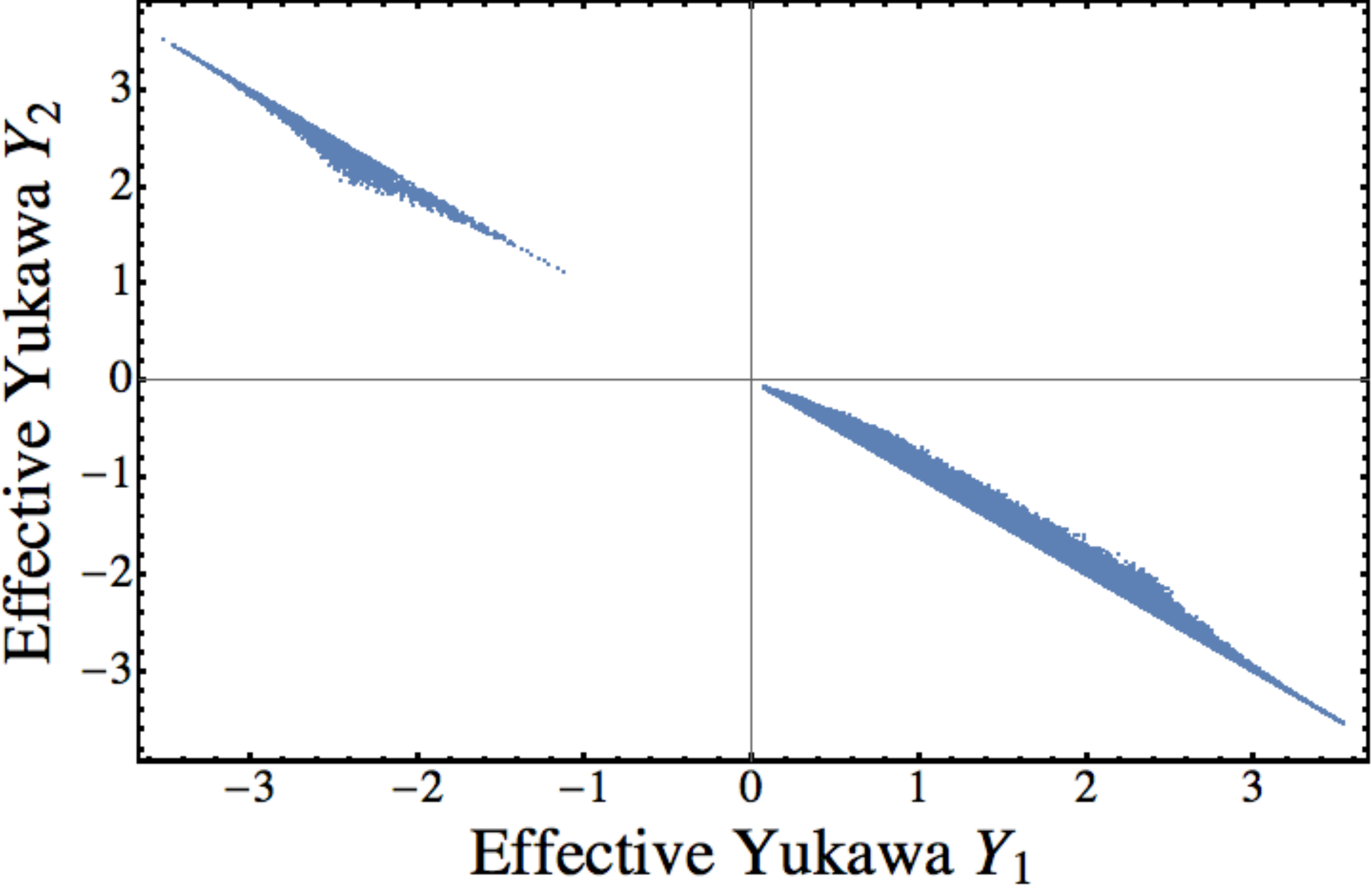}
  \caption{Regions of heavy Higgs cross-section as a function of the
    effective couplings $Y_1$ and $Y_2$ for octet VL fermions. The
    cross-section distribution over the plane is such that the larger
    the ``effective'' couplings the larger the cross-section.}
  \label{fig:sign-diff}
\end{figure}
For all points in the scan we have calculated (for each
representation) the exact value for $\alpha_{\rm s}$ making use of
eqs. (\ref{eq:beta}) and (\ref{eq:beta-1}). In doing so, we have
accounted for the decoupling of the different VL fermions at their
mass thresholds. For higher representations, in particular for
$\boldsymbol{10}$ and $\boldsymbol{15}$, a VL fermion mass way below
$m_H$ can lead to non-perturbative $\alpha_{\rm s}$ already at
$m_H$. Whenever those points are found we just drop them from our
analysis. If not stated otherwise, all our points are subject to the
cut $\Gamma_H<200\,$~GeV, where $\Gamma_H$ is the heavy Higgs total
decay width. Finally, for the calculation of the cross-section we have
used the MSTW PDFs at NNLO \cite{Martin:2009iq}.

The different ``effective'' couplings entering in (\ref{eq:x-sec}) are
weighted by different signs, with the sign difference holding
regardless of the region in parameter space, as shown in
fig.~\ref{fig:sign-diff}.  This effect leads to a certain degree of
cancellation of the different terms in the cross-section, something
that happens as well in the presence of color scalars as has been
pointed out in \cite{Dobrescu:2011aa}. As an illustration of this
cancellation, one can look at the particular case of
$y_U=\tilde{y_U}\equiv y$. In this case the mass matrix in
eq.~\ref{eq:up-type-quarks-mass-matrix} is symmetric, and
$\theta_R^U=\theta_L^U\equiv \theta$. Therefore, the effective Yukawas
entering eq.~(\ref{eq:Lag-up-type-nVLEq1}) read
$Y_{11}^U=-Y_{22}^U=y\,\sin (2\theta)/2$. One can then clearly see
that a cancellation in both, the up- and down-type sectors
contributions occurs up to mass non-degeneracy.

Several factors, however, ``compensate'' for such cancellation, and
can be sorted depending on whether they are or not
representation-dependent. Non representation-dependent correspond to
size of Yukawa couplings and VL fermion mass spectrum \footnote{Note,
  however, that the constraints on the mass spectrum are
  representation-dependent, and so indirectly it features a
  representation dependence.}. Group theory factors and $\alpha_S$
running are, instead, representation-dependent, and are such that for
representations beyond the triplet they lead to sizeable enhancements.
It is worth pointing out that for the fundamental representation, and
only for that representation, the ``compensating'' factors do not
suffice to render this possibility viable. However, if not for this
cancellation effect the fundamental color representation alone could
account for the diboson excess anomaly.

\begin{figure}
  \centering
  \includegraphics[scale=0.38]{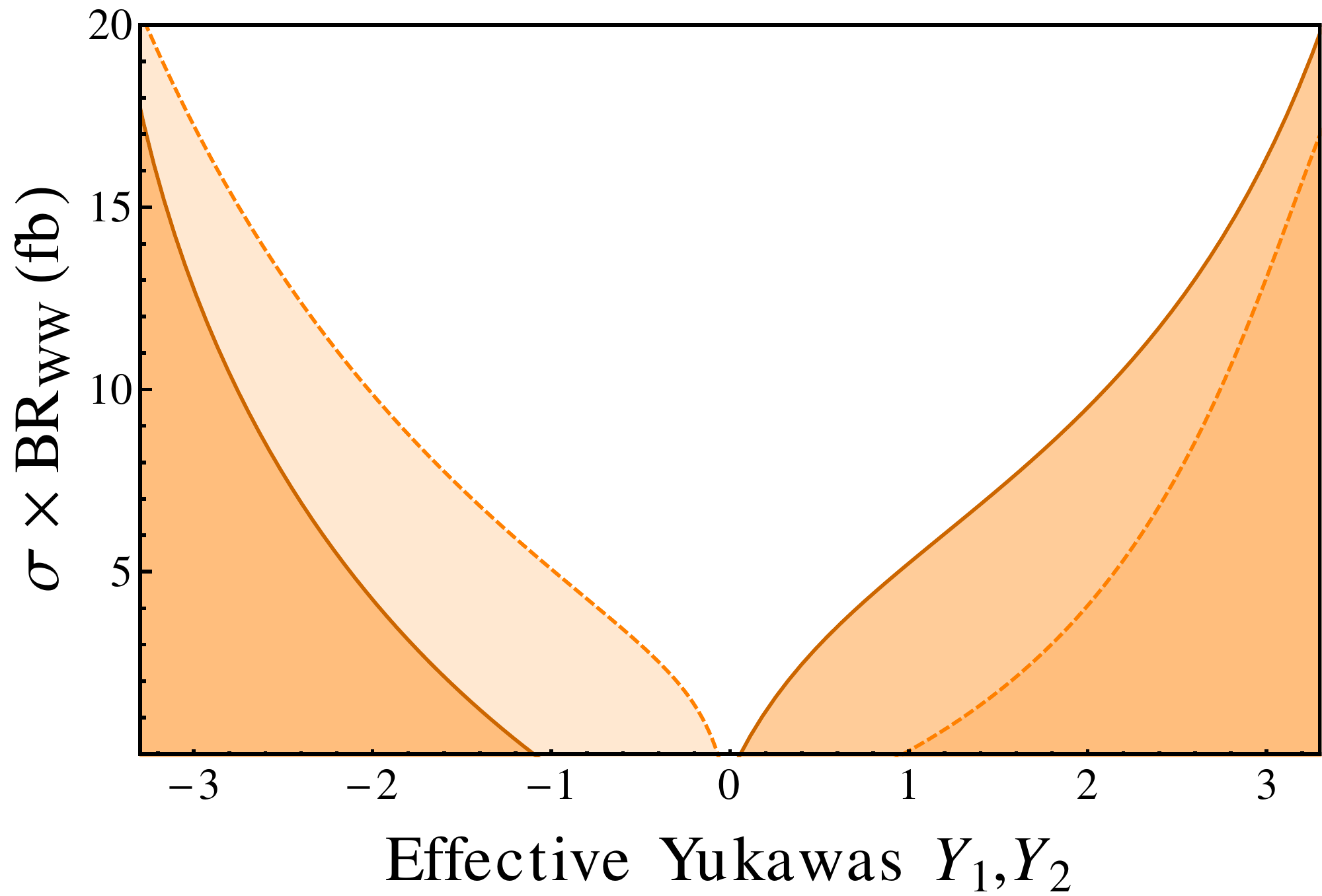}
  \caption{Signal $\sigma(pp\to H)\times \mbox{Br}(H\to W^+W^-)$ as a
    function of the effective couplings $Y_1$ (dashed orange) and
    $Y_2$ (solid brown) for octet VL fermions.}
  \label{fig:cs-vs-Yeffective}
\end{figure}
The larger the Yukawa couplings ($y$ and $\widetilde{y}$) the larger
the expected cross-section. We illustrate this in
figure~\ref{fig:cs-vs-Yeffective}, where we plot $\sigma(pp\to
H)\times \mbox{Br}(H\to W^+W^-)$ as a function of the effective
couplings $Y_1$ (dashed orange) and $Y_2$ (solid brown), for octet VL
fermions and $\Gamma_H\lesssim 200\,$~GeV. Note that the Yukawas are
typically $\mathcal{O}(1)$ at $m_H$ \footnote{One has to bear in mind
  that in models with higher $SU(3)_c$ representations $\kappa$ is
  expected to be larger than 2, and with larger $\kappa$ factors the
  Yukawa couplings will decrease accordingly.}. Thus, their RGE
running could lead in some cases to non-perturbative couplings or
vacuum instabilities at scales not-too-far from $m_H$, as it turns out
to be with $\alpha_S$. In that case, new degrees of freedom would be
needed to render our picture consistent at high energies.

\begin{figure}
  \centering
  \includegraphics[scale=0.2]{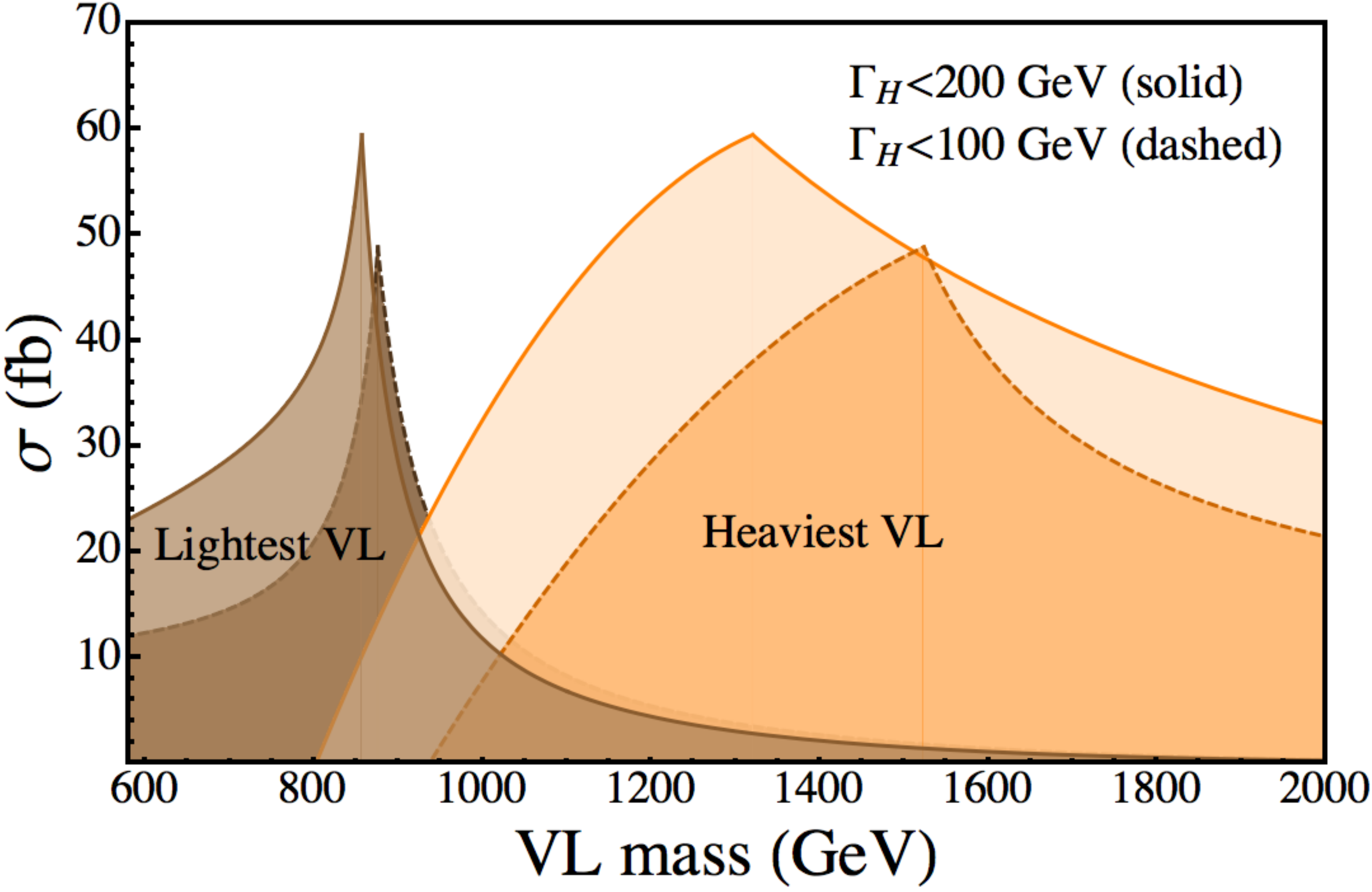}
  \caption{Gluon fusion cross-section versus the two octet VL fermion
    masses: the brownish region is for the lightest and the orangish
    for the heaviest. The different shaded regions correspond to the
    ``cuts'' $\Gamma_H<100$ GeV and $\Gamma_H<200$ GeV and show the
    constraints on the cross-section due to the condition of
    ``narrow'' resonance.}
  \label{fig:cs-vs-mvl-two-different-widths}
\end{figure}
However, large Yukawa couplings not only enhance the cross-section but
can potentially render the heavy Higgs total decay width well above
its maximum allowed value, $\Gamma_H\sim 200\,$~GeV.  The ``narrow''
width condition places a strong constraint on the possible values of
the Yukawa couplings, with the effect being more pronounced for higher
representations.  The reason is rather simple. While only two Yukawa
vertices contribute to the gluon fusion cross-section (first and
fourth terms in ecs. (\ref{eq:Lag-up-type-nVLEq1}) and
(\ref{eq:Lag-down-type-nVLEq1}), for $A=2$), four contribute to
$\Gamma_H$, determined by partial decay widths weighted by final state
multiplicities, whose values scale with the dimension of the
representation. In fig. \ref{fig:cs-vs-mvl-two-different-widths}, we
display results for the gluon fusion cross-section as a function of
the two octet VL fermion masses: the brownish region is for the lightest
and the orangish for the heaviest. The results correspond to two
different ``cuts'', determined by $\Gamma_H<100\,$ GeV and
$\Gamma_H<200\,$ GeV. This shows that indeed the smaller $\Gamma_H$,
the smaller the cross-section. Furthermore, the role of kinematically
open/closed Higgs fermion channels is also striking. In those regions
of ``light'' states ($m_{X_1}<m_H/2$) the cross-section is small and
increases towards values approaching the fermion modes kinematical
threshold, reaching a maximum determined by the loop function
$A(m_H^2/m_{X}^2)$ and decreasing due to the expected decoupling
behavior of the cross-section.

Representation-dependent effects are obvious but remarkable. As shown
in tab. \ref{tab:casimir}, the heavy gluon fusion cross-section
rapidly increases with higher representations. Thus, alone, the group
theory factor could ``overcome'' the cancellation effect.  However,
higher representations in turn are, in the case of prompt decays,
subject to more stringent experimental constraints, thus being more
sensitive to the cross-section decoupling behavior. With the bounds we
derived in sec. \ref{sec:pheno-analy} for short-lived VL fermions, we
find that while the sextet and octet lead to a signal, $\sigma(pp\to
H)\times \mbox{Br}(H\to W^+W^-)$, in agreement with the ATLAS reported
excess, $\boldsymbol{10}$ and $\boldsymbol{15}$ do not. Such statement
is only valid when those mass bounds hold, deviations from those
values will of course change the conclusion. For example, if these
representations are long-lived their mass limits will not be so
stringent, allowing them to perfectly fit the ATLAS anomaly (see the
discussion in sec. \ref{sec:pheno-analy}). For that reason, we do not
discard the possibility of $\boldsymbol{10}$ and $\boldsymbol{15}$ VL
fermions {\it a priori} and calculate the signal for all the
representations listed in tab. \ref{tab:casimir}.

\begin{figure}
  \centering
  \includegraphics
  [scale=0.2]{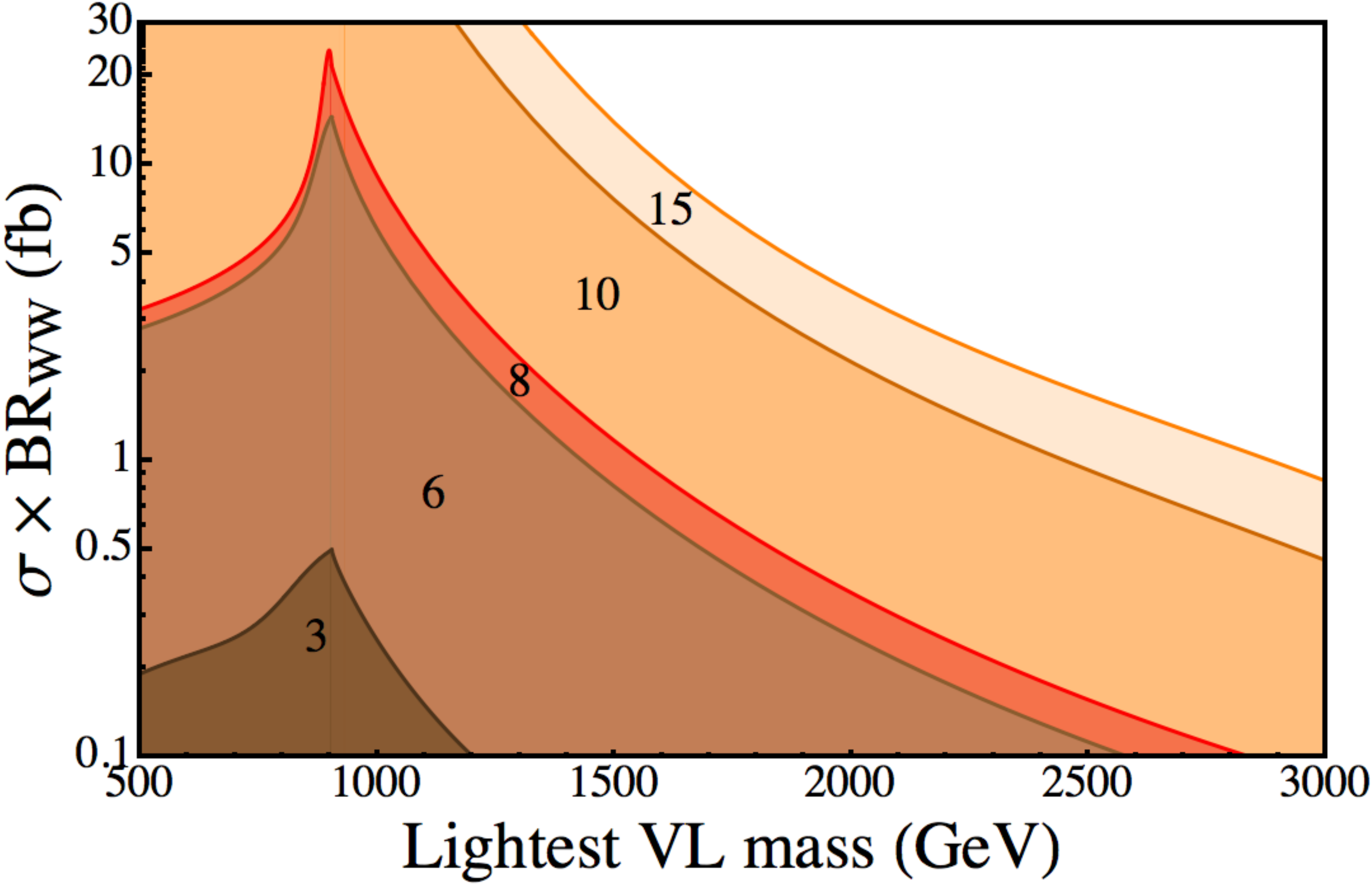}
  %scan_with_running_final_morecolors.pdf
  \caption{Signal $\sigma(pp\to H)\times \mbox{Br}(H\to W^+W^-)$ as a
    function of the lightest VL fermion for the five color
    representations we have employed. The different shaded regions
    from bottom to top correspond to: $\boldsymbol{3}, \boldsymbol{6},
    \boldsymbol{8}, \boldsymbol{10}, \boldsymbol{15}$.}
  \label{fig:signal-vs-mVL}
\end{figure}
Fig. \ref{fig:signal-vs-mVL} shows the results for the five
representations we are considering. There relative values agree very
well with those expected from group theory, see
table~\ref{tab:casimir}. This result clearly shows that the
fundamental representation cannot, by any means, account for the ATLAS
excess.  Even relying on the mass bounds derived for short-lived
fermions, the sextet and octet can readily address the ATLAS
observation provided $m_{X_{6,8}}\lesssim 1.5$ TeV. The
$\boldsymbol{10}$ and $\boldsymbol{15}$ can account for the anomaly
depending on their mass limits.\footnote{Note that even using the most
  stringent bounds they can yield a consistent signal in the case
  $n_\text{VL}>1$.} For masses of up to about $\sim 2.5$ TeV, both
representations can fit the observed ATLAS signal (with a somewhat
marginal fit at 2.5 TeV). For masses above those values the decoupling
effect is stringent and the signal is degraded below 1 fb, way below
the values indicated by ATLAS. For those representations too, one may
wonder about the extremely ``large'' signal at low VL fermion masses:
such values can be fitted to those required to address the anomaly by
properly decreasing the values of the Yukawa couplings.

\begin{figure}
  \centering
  \includegraphics
  [scale=0.2]{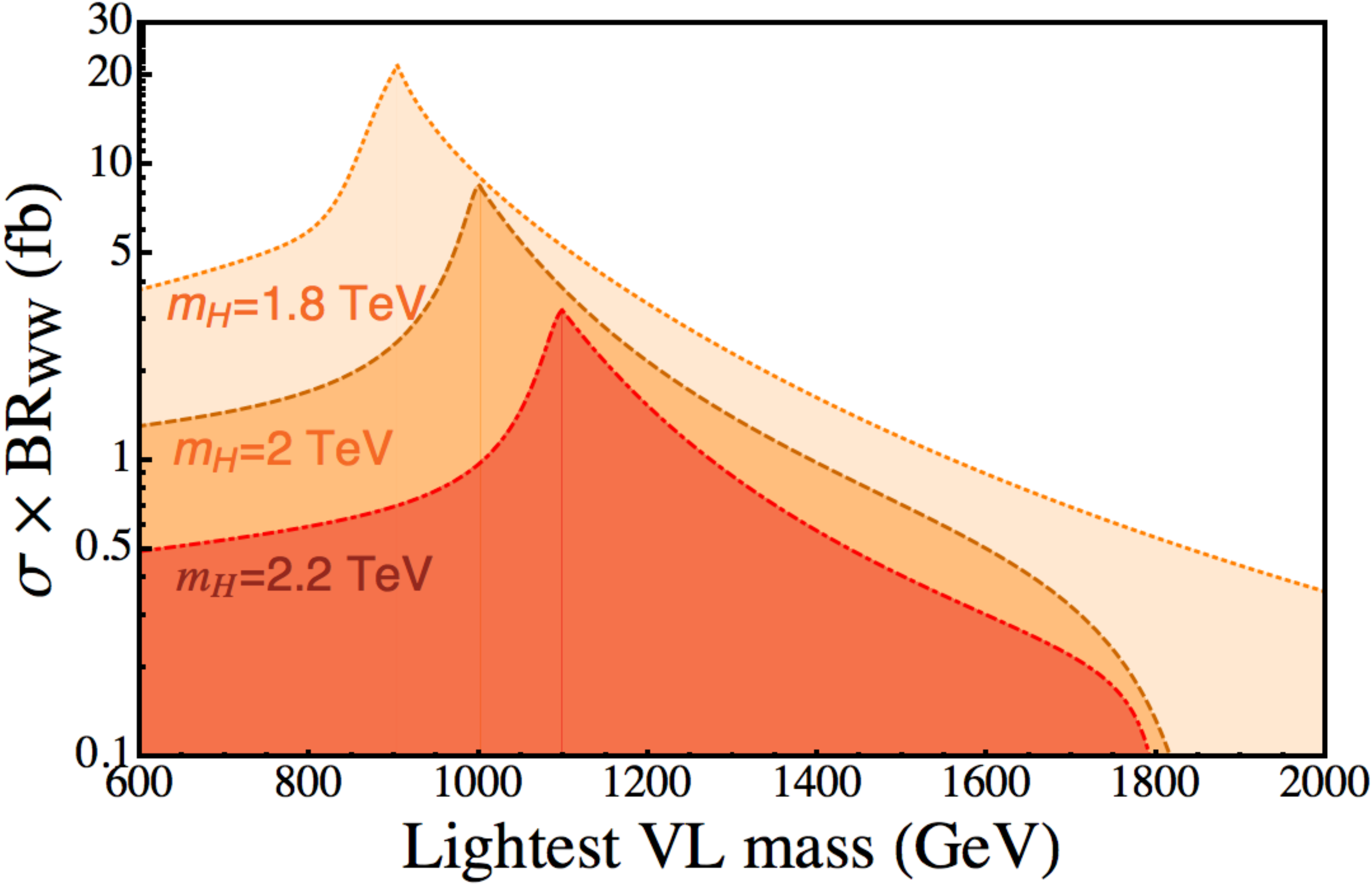}
  %sigmaBR_mHs.pdf
  \caption{Signal $\sigma(pp\to H)\times \mbox{Br}(H\to W^+W^-)$ as a
    function of the lightest VL fermion for the octet. The different
    shaded regions from top to bottom correspond to: $m_H=1.8, \,2$
    and $2.2$ TeV.}
  \label{fig:signal-vs-mVL-for-diff-mH}
\end{figure}

The heavy Higgs gluon fusion cross-section dependence on $m_H$ is as
well somewhat strong. Thus, since the results presented so far are for
$m_H=1.8\,$~TeV, we have investigated up to which extent the octet and
sextet can or not account for the signal in the relevant experimental
range, [1.8, 2.2] TeV. Note that in the low mass region the
$\boldsymbol{10}$ and $\boldsymbol{15}$ are expected to always be able
to address the anomaly, regardless of the Higgs
mass. Fig. \ref{fig:signal-vs-mVL-for-diff-mH} shows the results for
the octet case for three different values of $m_H: 1.8, 2.0,
2.2\,$~TeV. Although the signal is depleted about an order of
magnitude when moving from 1.8 TeV to 2.2 TeV, it is still possible to
obtain a signal within the range reported by ATLAS.

\section{Conclusions}
\label{sec:conclusions}
% \draftnote{With Juan we have realized several of those points raised
%   in ``Discussion'' are not worth mentioning. Keeping discussion +
%   conclusions seems not appriopriate then. I'll then extend a bit more
%   this section and suppress the ``Discussion'' section. I've not 
%   edited this part whatsoever!}\\
We have shown that the diboson excess reported by ATLAS (and CMS)
might be due to the production and further decay of a heavy Higgs,
$H$, resulting from a type-I 2HDM. Production proceeds through gluon
fusion, enhanced by the presence of colored VL fermion. In addition to
``standard'' color triplets, we have considered as well higher-order
color representations (QCD exotics) which we have taken to be
$\boldsymbol{6}$, $\boldsymbol{8}$, $\boldsymbol{10}$ and
$\boldsymbol{15}$. Our findings show that barring the triplet case (in
its minimal form), all other representations lead to large
cross-sections in fairly large portions of the parameter space.

We have studied constraints on VL fermion masses, which we have argued
depend upon their lifetime. However, no matter whether the new states
are short- or long-lived we have found that---in
general---phenomenological consistency requires their masses to be
above 1 TeV. These limits then translates into heavy Higgs decays
dominated by gauge boson modes, thus naturally yielding the
$W^\pm\,W^\mp$ and $Z^0\,Z^0$ diboson signal observed by ATLAS.

The different scenarios we have considered can be regarded as
minimal. Additional VL fermion generations could be considered as
well, and in those cases enhancements of the cross-section are
expected. These non-minimal scenarios are of particular interest in
those cases where constraints on the VL fermion masses are
stringent. As we have demonstrated, decoupling in such cases is severe
and strongly depletes the cross-section. Thus, the inclusion of
additional generations can potentially open regions of parameter space
that otherwise are closed.

In addition to addressing the results reported by ATLAS, the heavy
Higgs resonance we have put forward leads to several other
\textit{remarkable predictions}. First of all, the to-some-extent
small mass splitting between the heavy CP-even and CP-odd states leads
necessarily to triboson signatures, $W^\pm\,W^\mp\,Z$ and $Z\,Z\,Z$
\cite{Aguilar-Saavedra:2015rna}. Secondly, charged Higgs single
production, being Cabibbo suppressed and driven by SM couplings is
negligible. Thus, it is not possible to generate an excess in the $W^\pm\,Z$ channel in this setup. Finally, sufficiently large $H$ cross-sections require
vector-like fermion masses below $\sim 3\,$~TeV, hence being
potentially producible at LHC.

We conclude by emphasizing that if the $\sim2$ diboson resonance were
to be confirmed, LHC data should as well tell us soon whether the
mechanism we have pointed out here is responsible for such
observation: the diboson signal should be accompanied by a CP-odd
Higgs whose mass should not exceed $\sim 2.3\,$~TeV, TeV-colored
fermions should be copiously produced, no statistically significant
diboson events in the $W^\pm\,Z$ channel should be observed, while there should be a signal in the triboson channels ($W^\pm\,W^\mp\,Z, Z\,Z\,Z$). Therefore, the model we have discussed here will be soon subject to deep experimental scrutiny.

\section*{Acknowledgements}
We would like to thank Frank Deppisch, Martin Hirsch, Suchita Kulkarni
and Werner Porod for useful conversations.  DAS would like to
acknowledge financial support from the Belgian FNRS agency through a
``Charg\'e de Recherche'' contract.
%
%BIBTEX:
%pdflatex Atlas-excess
%bibtex Atlas-excess.tex
%pdflatex Atlas-excess
%pdflatex Atlas-excess
\bibliographystyle{h-physrev5}
\bibliography{refs}
\end{document}